# Full-field Brillouin microscopy based on an imaging Fourier transform spectrometer

Carlo Bevilacqua[1] and Robert Prevedel[1-4]

[1] Cell Biology and Biophysics Unit, European Molecular Biology Laboratory, Heidelberg, Germany.
[2] Developmental Biology Unit, European Molecular Biology Laboratory, Heidelberg, Germany
[3] Epigenetics and Neurobiology Unit, European Molecular Biology Laboratory, Rome, Italy.
[4] German Center for Lung Research (DZL), Heidelberg, Germany.

Correspondence should be addressed to R.P. (prevedel@embl.de).

**Brillouin microscopy is an emerging optical elastography technique that can be used to assess mechanical properties of biological samples in a 3D, all-optical and hence non-contact fashion. However, the low cross-section of spontaneous Brillouin scattering results in weak signals typically requiring prolonged exposure times or illumination dosages potentially harmful for biological samples. Here, we present a new approach for highly-multiplexed, and therefore rapid, spectral acquisition of the Brillouin scattered light. Specifically, by exploiting a custom-built Fourier-transform imaging spectrometer and the symmetric properties of the Brillouin spectrum, we experimentally demonstrate full-field 2D spectral Brillouin imaging of phantoms as well as biological samples, at a throughput of up to 40,000 spectra per second over a ~300μm field-of-view. This represents an approximately three orders of magnitude improvement in speed and throughput compared to standard confocal methods while retaining high spatial resolution and the capability to acquire three-dimensional images of photosensitive samples in biology and medicine.**

**Introduction**

Mechanical properties of cells and tissues such as elasticity and viscosity are important parameters that have been shown to play crucial roles in determining biological function[1,2], however standard techniques currently used to assess them exhibit intrinsic limitations. Among them, atomic force microscopy (AFM) or micropipette aspiration, the gold standard techniques currently used in mechanobiology fields, both require contact forces and are intrinsically restricted to measurements at a sample's surface. Other optical approaches, such as optical coherence elastography (OCE), requires external contact forces or ultrasound fields to measure tissue displacements. While OCE enables rapid three-dimensional imaging, current implementations do not achieve cellular resolution, a limitation shared with other non-invasive techniques, such as ultrasound and magnetic resonance imaging.

In recent years, Brillouin microscopy[3,4], a type of optical elastography, has emerged as a non-destructive, label- and contact-free method that can probe the viscoelastic properties of biological samples with diffraction-limited resolution in 3D. It thus offers a conceptually novel way to probe elastic and viscous properties of biomaterials[5,6]. Brillouin light scattering is an inelastic process arising from the interaction of light with spontaneous, thermally induced density fluctuations, so called acoustic phonons. This interaction gives raise to two additional peaks in the scattered light spectrum known as Stokes and Anti-Stokes Brillouin peaks (see **Fig. 1a**). Analysis of the Brillouin spectrum can then provide, for a known material density and refractive index, a unique characterization of the material's mechanical properties. In particular, the Brillouin spectral shift (peak position) is related to the elastic properties (longitudinal modulus) of a material, while its linewidth is related to its (longitudinal) viscosity.

Regrettably, the spontaneous Brillouin scattering cross-section is weak, resulting in low scattering probabilities ($10^{-12}$) that consequently necessitate long signal integration times and thus result in very slow imaging speeds with typical measurement times of tens to hundreds of milliseconds for a single datapoint (i.e. minutes to hours for 2D images of 50-250px$^2$)[7–9]. While recent work has made substantial progress by either multiplexing the data acquisition by illuminating and collecting spectra along an entire line[10–12] or relying on stimulated Brillouin scattering approaches[13–15], overall imaging speed performances has remained far from fluorescence microscopy modalities which lies in the range of μs per pixel. In particular, despite promising recent work in this direction[16,17], there currently does not exist a practical solution for two-dimensional (2D) multiplexing akin to light-sheet microscopy that allows capturing entire 2D Brillouin images simultaneously.

In this work, we present a new approach for Brillouin spectroscopy based on a custom Fourier-transform (FT) imaging spectrometer and an optimized sampling approach specifically designed for the high, sub-picometer (i.e., sub-GHz) spectral resolution required for Brillouin imaging. Importantly, the use of an imaging spectrometer enables 'full-field' measurements of entire 2D planes concurrently with array detectors (cameras), similar to a wide-field or light-sheet[18] microscope. While FT imaging spectroscopy has found numerous applications in fields such as general hyperspectral imaging[19,20] or Raman scattering based microscopy[21], applications to Brillouin spectroscopy have so far been challenging due to the required high spectral resolution (sub-GHz), which in turn necessitates to acquire a large number of samples, over a large range of interferometer delay positions (typically ~$10^6$). Key to our approach is the realization that the symmetric properties of the Brillouin spectrum can be exploited to substantially (by more than 10.000-fold) reduce the number of measurements to reconstruct a typical Brillouin spectrum. When combined with narrowband spectral filtering via atomic gas cells[22], this allows to suppress the dominating Rayleigh background which in turn enables practical Brillouin imaging applications in biology and beyond. Overall, our approach – termed Fourier-transform Brillouin microscopy (FTBM) - provides a substantial speed-up for a 2D Brillouin spectral imaging, as conventional cameras can record millions of pixels simultaneously, which leads to a substantial reduction in terms of overall imaging time, as well as light exposure, which is highly advantageous for a number of (biological) imaging applications. We demonstrate the effectiveness and capabilities of our FTBM to record Brillouin microscopic imagery at high spatial and temporal resolution by imaging heterogenous phantoms as well as live zebrafish larvae.

**Results**

*Principle of the FTBM approach*

The principles underlying Brillouin scattering based microscopy, as well as our particular approach, are shown in **Fig. 1**. FT spectroscopy is an established technique for characterizing the spectral properties of a light source by measuring an interference signal in the time-domain. Typically, this is achieved by measuring the optical power at the output of a (Michelson) interferometer as a function of the optical delay, given by the difference in interferometer arm length. Applying a Fourier transform to the acquired data then allows to retrieve the power spectral density as a function of optical frequency or wavelength.

Specifically, in case of a Michelson interferometer (**Fig. 1b**), the intensity at the output of the interferometer is given by:

$$I(\tau) = \langle |E(t)/\sqrt{2} + E(t-\tau)/\sqrt{2}|^2 \rangle_t = I_0 + \langle E(t)E(t-\tau) \rangle_t$$

where $E(t)$ is the optical electric field, $\langle \cdot \rangle_t$ indicates the average over the integration time of the detector, $\tau = \frac{2\Delta L}{c}$ is the time delay between the two arms of the Michelson interferometer and $I_0 = \langle |E(t)|^2 \rangle_t$.

The term $\langle E(t)E(t-\tau) \rangle_t$ is the autocorrelation function for the electric field and, according to the Wiener–Khinchin theorem, corresponds to the Fourier transform of the optical spectrum. Hence, FT spectroscopy determines the optical spectrum from the interferogram $I(\tau)$. Practically, one samples $I(\tau)$ at discrete intervals, $\tau_n = \frac{2\Delta L_n}{c}$. The sampling rate and the number of samples required to properly reconstruct the spectrum from $I(\tau)$ can be easily found by considering the Nyquist–Shannon sampling theorem and the properties of the discrete Fourier transform. Specifically, the required sampling rate is at least twice the maximum optical frequency present in the optical spectrum, i.e. $\delta\tau_n < \frac{\lambda_{min}}{2c} \rightarrow \delta\Delta L_n < \frac{\lambda_{min}}{4}$. In turn, the maximum time delay $\tau_{max}$ determines the spectral resolution $\Delta\nu$ achievable (**Fig. 1c**). The required number of samples thus becomes:

$$N = \frac{\tau_{max}}{\delta\tau_n} > \frac{1/\Delta\nu}{\lambda_{min}/(2c)} = \frac{2c}{\Delta\nu \cdot \lambda_{min}} \quad (1)$$

For measurements of a typical Brillouin spectrum, e.g. at a wavelength of $\lambda_{min}$=780nm, for which a high spectral resolution of $\Delta\nu$=0.5GHz is normally required, the overall number of samples becomes $N>10^6$. If we consider 10-100ms integration time, which is a typical value for acquiring Brillouin spectra from biological samples[23], this practically will require between ~3-30 hours. Prior work has explored the use of compressive sensing approaches to reduce the number of required samples in Raman FT spectroscopy, and has found an improvement by ~4-folds by exploiting the sparsity of the Raman spectrum[24]. In contrast, in our work we show that, by exploiting the symmetry of the Brillouin spectrum, we can reduce the required number of samples by >$10^4$ folds, as outlined briefly below (for more details, see **Supplementary Note 1**)

The main realization is that for a symmetric band-limited spectrum, the interferogram $I(\tau)$ consists of a fast-oscillating term at the central frequency enclosed by a slowly varying envelope which contains the full the spectral information. Thus, the optical spectrum can be measured by locally determining the envelope of the interferogram. This can be achieved by sampling the interferogram at the central frequency (laser optical frequency, $\omega_L$) with only few points $N_L$ (in principle $N_L = 3$ points are sufficient if only the amplitude, phase and offset need to be recovered), from which the signed amplitude of the oscillation $A(\tau)$ can be determined (**Fig. 1d** - for additional details on how to determine the sign, see **Supplementary Note 2**). Consequently, we only need to apply the sampling conditions to the function $A(\tau)$ instead of to the full interferogram $I(\tau)$. Therefore, depending on the spectral components in $A(\tau)$, the interferogram can be largely undersampled while still retaining the complete spectral information. In particular if the bandwidth of the spectrum is upper-limited by some $\omega_{max} + \Omega$, then $A(\tau)$ can be reconstructed by sampling it at $\delta\tau_n < \frac{\pi}{\omega_{max}+\Omega}$.

$$N = N_L \frac{\tau_{max}}{\delta\tau_n} > N_L \frac{1/\Delta\nu}{\pi/(\omega_{max} + \Omega)} = N_L \frac{\omega_{max} + \Omega}{\pi\Delta\nu} \quad (2)$$

For Brillouin spectra measured from typical biological samples at 780nm, $\omega_{max} + \Omega \lesssim 2\pi \cdot 7GHz$, and therefore $N > N_L \frac{7GHz}{\Delta\nu} \approx 30$, which is an at least $10^4$ folds reduction compared to the standard approach requiring $N>10^6$ samples, as calculated from Eq. (1).
We note that our derivation is not limited to Brillouin spectroscopy and can be applied to any symmetric spectrum. In general, the reduction in the number of required points $r$ is given by the ratio of eq. (1) and eq. (2):

$$r = \frac{2c}{\Delta\nu \cdot \lambda_{min}} \frac{1}{\left(N_L \frac{\omega_{max} + \Omega}{\pi\Delta\nu}\right)} = \frac{2\pi c}{\lambda_{min} N_L (\omega_{max} + \Omega)} \approx \frac{\omega_L}{N_L (\omega_{max} + \Omega)} \quad (3)$$

which corresponds to the ratio between the optical frequency, $\omega_L$, and the frequency range $\omega_{max} + \Omega$ over which one aims to reconstruct the spectrum.

Finally, we note that the finite linewidth of the Brillouin peaks $\Delta\nu_B$ causes an exponential decay of the amplitude, thus effectively setting an upper limit to the optical path difference which makes sense to measure (see Eq. S1.5).

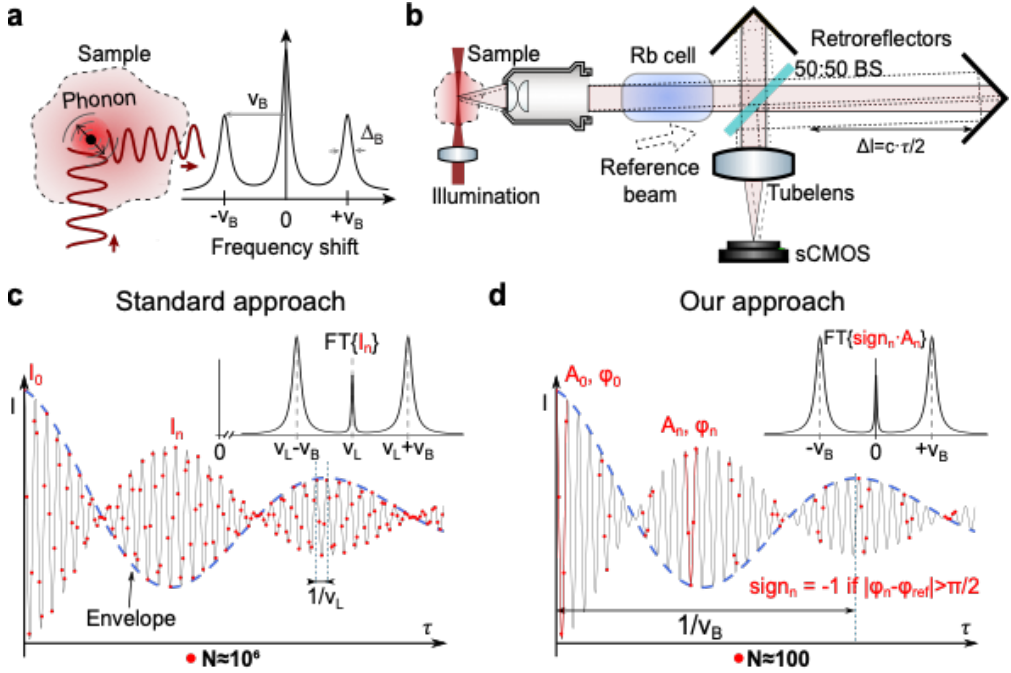

**Figure 1: Principle of FT-Brillouin microscopy (FTBM). a**, In Brillouin scattering a small portion of the incident light (dark red, upwards pointing arrow) interacts with thermal phonons intrinsically present in the sample. This gives rise to a scattered light (Brillouin) spectrum with two symmetrical peaks on the side of the main laser line with shift and width as indicated. **b**, Conceptual schematic of the FTBM: the sample is illuminated with a light-sheet (side-view only here); an objective lens collects the scattered light and a Rubidium (Rb) cell suppresses the elastically scattered light without affecting the Brillouin signal. After going through a Michelson interferometer, a tube lens forms the image of the sample on the sensor; a reference beam is also introduced to provide a reference for determining the optical phase (see **Supplementary Note 2**). **c,d**, Comparison of sampling the interferogram (i.e. the intensity on the detector vs. the time delay $\tau$ between the two arms of the Michelson) between "standard" FT spectroscopy (c) and our approach (d). In the standard approach the full interferogram is sampled according to the Nyquist–Shannon criterion, requiring $\approx 10^6$ points (red dots) to achieve the necessary spectral resolution of ~500MHz. **d**, In our approach the interferogram is only locally sampled, with much fewer points, to reconstruct the amplitude (blue dashed line) of the envelope, while the local phase is used to determine its sign; our approach requires only $\approx 100$ points. Insets: In both cases the full spectrum can be reconstructed by performing the FT of the samples, but in the case of the standard approach the spectrum is centered at the laser frequency $\nu_L$ while in our approach it is centered at 0.

*Experimental performance characterization*
To demonstrate the feasibility of our sub-sampling approach for experimental Brillouin measurements, we designed and built a custom FT-interferometer and coupled it to an inverted SPIM microscope (adapted from[10], see Methods for more details). The FT-interferometer consists of a Michelson interferometer, placed in the infinity-space of the imaging system (**Fig. 1b**), and uses corner-cube reflectors for high stability of operation[25]. A Rubidium-87 vapor cell, placed before the Michelson, is used to suppress the elastically scattered light while not affecting the Brillouin signal[10]. A reference beam, taken from the laser, is also introduced in the interferometer to aid in determining the sign of the amplitude (see Methods and

**Supplementary Note 2** for more details). A detailed optical schematic of the setup can be found in **Extended Data Fig. 1**.

To characterize the performance of our FTBM system, we imaged a heterogeneous phantom consisting of oil beads embedded in agar over a ~300μm wide field of view (**Fig. 2, Supplementary Video 1**). The Brillouin shift and width for each point are extracted by fitting the interferogram with the function S1.5 (derived in **Supplementary Note 1**). An example interferogram with the corresponding fit is shown in **Fig. 2b**. The FT of the interferogram is also shown to highlight the quality of the optical spectrum that can be recovered from it. The spatial heterogeneity of the sample, especially the presence of agar inclusions inside the oil (see inset of **Fig. 2a**) allows to showcase the high spatial and mechanical resolution of our FTBM system on micrometer length scales. Specifically, we quantified the spatial resolution by evaluating a sharp transition between agar and oil at the edge of a bead. To this end we performed a double peak fit on the spectra along the transition and plotted the amplitude of the oil and agar component (**Fig 2c**). A fit with an *erf* function shows a FWHM of ~1.2μm consistently for the two curves. **Fig. 2d** shows the spectra along a different (less sharp) transition where the two peaks (from oil and agar) with a varying amplitude are clearly visible. To further corroborate the high imaging resolution and degree of spectrometer alignment, we placed an USAF resolution target in the intermediate image plane (as indicated in **Extended Data Fig. 1**) and imaged it through the Michelson at different position of the scanning arm, while blocking the reference arm. This demonstrates that the resolution and magnification of the optical system can be preserved even with an increase in optical path of $2 \cdot \Delta l = 600$mm (**Extended Data Fig. 2**).

The throughput of our approach can be calculated by considering that a single 2D image from the stack shown in **Fig. 2a** consists of 621x735 px$^2$ (spectra), reconstructed from 155 individual scanning arm positions (31 coarse and 5 fine steps), acquired by 155 camera images with 100ms exposure time for each, for a total data integration time of 15.5s (see Methods). This corresponds to a throughput of ~30k spectra/s. We note that in our current implementation the actual acquisition time is slightly longer, due to stage movement, but it can straightforwardly be improved by using a faster (direct drive) stage. Finally, to characterize the spectral precision, we generated a histogram for the Brillouin shift and width from a large homogeneous oil region (**Fig. 2e**), and found a standard deviation of 78MHz and 270MHz, respectively, as expected from theoretical simulation (see next section). To show that the shift precision is homogeneous across the entire FOV we plotted a spatial map of the precision for water (**Extended Data Fig. 3**), and obtained shift and width precisions of 83MHz and 240MHz, respectively, consistent with the results obtained for oil in the heterogenous phantoms.

We highlight that in principle, the same setup and approach could be used for Raman imaging, providing complementary chemical information on the sample, which can aid in further interpretation of the measured mechanical properties[26,27].

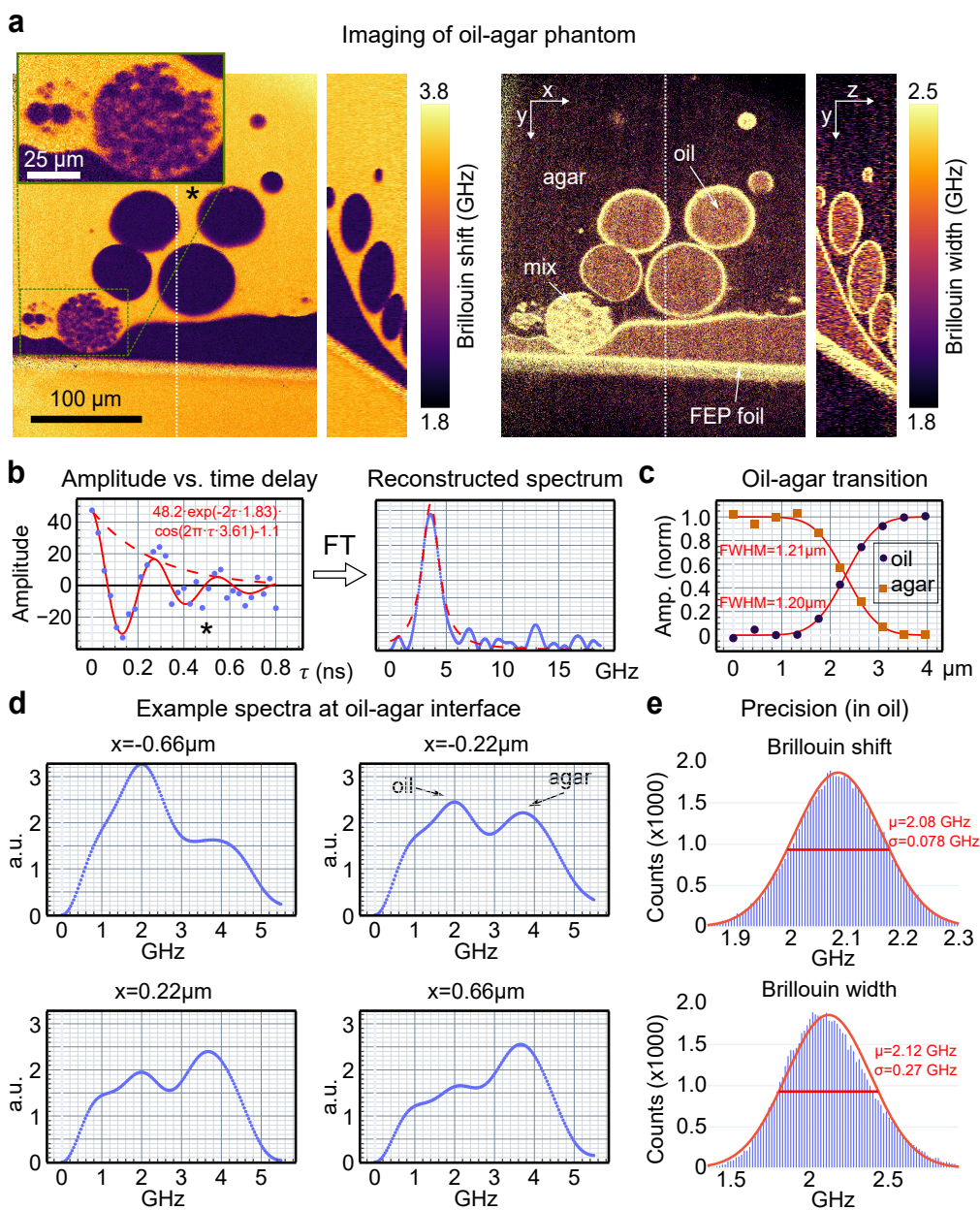

**Figure 2: Experimental characterization of the FTBM: a**, 3D imaging from a material phantom, consisting of oil droplets in agar (see Methods), both for the Brillouin shift (left) and Brillouin linewidth (right); the FOV of a single plane is 273x323µm, corresponding to 621x735 pixels; the white vertical dotted line indicates the plane shown in the orthogonal view; the asterisk '*' indicate the position of the spectrum shown in panel b. Also see **Supplementary Video 1**. **b**, Example interferogram (left) and corresponding Fourier Transform (FT, right). Note that the absolute value of the plotted amplitude corresponds to actual number of photoelectrons, as detected by the camera. Solid red line in the left panel shows the fit to experimental interferogram data whose parameters are stated in red text while the dashed red line indicates the exponential decay. The dashed red line in the right panel indicates the Fourier Transform of the function on the left (and not a fit performed on the spectrum). **c**, Plot of the amplitude of the "oil component" (at 2.10GHz) and the "agar component" (at 3.49GHz) through the edge of a large bead (taken from the stack shown in panel a). Each point is calculated by fitting the spectrum (taken as the average of 9 adjacent points) with two Gaussian peaks centered at the oil and agar Brillouin shift

respectively. The solid lines show a fit with an *erf* function, from which the FWHM is derived. **d**, Example spectra (average of 6 adjacent points) along the transition between oil and agar, showing the two peaks at the oil and agar Brillouin shift. Note that it is a different transition than the one showed in panel c. **e**, Histogram representing the precision for Brillouin shift (top) and width (bottom) for a large oil region taken from the stack shown in panel a; the solid red curve represents a Gaussian with the mean and standard deviation calculated from the data and as indicated in red text.

*Numerical performance evaluation*

Next, we sought to compare our experimental results with theory, in order to evaluate how the different experimental parameters might affect the performance of our measurements. To this end, we devised the following numerical simulation framework: We assumed a typical experimental Brillouin spectrum of water (Lorentzian shape, shift 3.49GHz, FWHM 1.6GHz at 780nm), and computed the FT-interferometer transfer function (see **Supplementary Note 1** for derivation, Eq. S1.5) in terms of the number of detected photo(electro)ns on the camera $N_{detect}$ as a function of the optical delay, $\tau$:

$$2N_{detect} = N_{ASE} + N_{elas} + N_{Br} + \left[N_{elas} + N_{Br} \cdot e^{-2|\tau| \cdot \Delta \nu_B} \cdot \cos(\Omega \tau)\right] \cdot \cos(\omega_L \tau) \quad (4)$$

where $N_{ASE}$ is the number of photons for the Amplified Spontaneous Emission, $N_{elas}$ is the number of elastically scattered photons and $N_{Br}$ is the number of Brillouin scattered photons. To make these simulations as realistic as possible, we further considered various sources of experimental noise and imperfections, such as shot noise, camera noise, limited stage precision and intensity noise. We then locally sampled this interferogram with n = 5 samples, spaced by a change in optical path length (OPL) of $\lambda/4$=195nm, and reconstructed the local amplitude and phase by fitting a cosine function with fixed frequency (given by $4\pi/\lambda$). The same procedure is repeated after increasing the OPL iteratively by 20 larger steps of 10mm to estimate the overall interferogram's envelope. We note that OPL=2$\Delta L$, which is important when comparing the simulation with the experimental data. The Brillouin shift and linewidth are then determined by least-square fitting of the so obtained datapoints with Eq. S1.5.

Our numerical simulations in general agree well with experimental measurements, both showing the expected shot-noise limited performance (log-log slope of –0.5, see **Fig. 3**). Furthermore, we found that a high level of precision in the Brillouin shift (<20MHz) and width (<100MHz) estimates can be obtained with only ~1000 detected photons when scanning over 20 discrete OPL steps (**Fig. 3a,b**). Importantly, our simulations provide valuable scaling laws of the expected performance of our FTBM system with respect to the detected signal intensity as well as various sources of experimental error, including camera read-out noise (**Fig. 3a,b**), stage precision (**Fig. 3c**), as well as Rayleigh, ASE and intensity noise (**Fig. 3d,e**). Note that practically ASE intensity levels can be highly dependent on the sample and its amount of elastic scattering. Finally, the simulations provide guidance towards the expected precision as a function of the OPL sampling steps and overall signal level (**Fig. 3f**). Overall, the results show that state-of-the-art Brillouin measurements (precision <10MHz) can be expected even with consumer grade cameras or translation stages and assuming realistic Brillouin signal levels.

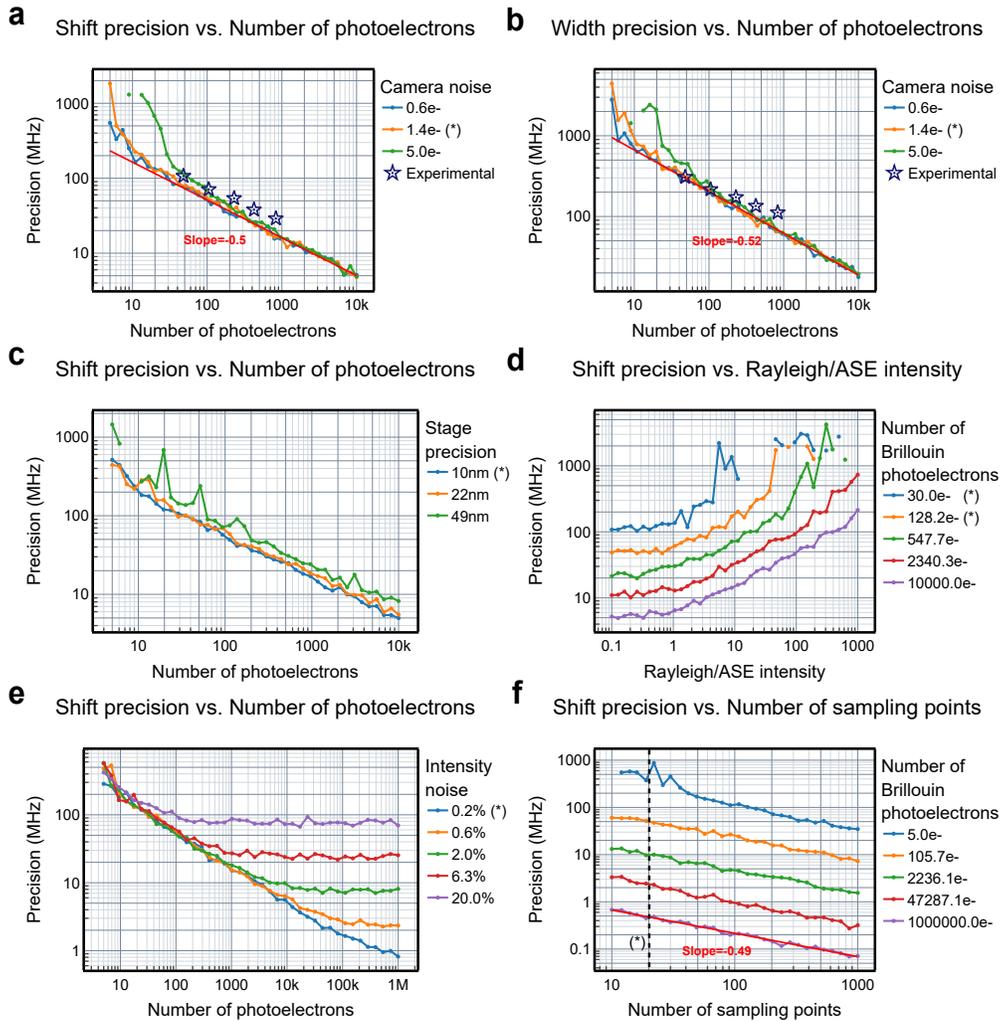

**Figure 3: Numerical simulation of the FTBM performance. a,b**, Shift and width precision as a function of the number of detected photons at varying readout noise levels of the camera (reported as standard deviation). The stars indicate experimentally measured values in water (see Methods) and confirm the quantitative validity of our simulations. **c**, Shift precision as a function of the number of detected photons at varying FT-interferometer stage precision (reported as standard deviation). **d**, Shift precision as a function of Rayleigh intensity at varying number of detected photons. Note that the effect of the ASE intensity on the precision is similar which is why we plot them on the same scale. **e**, Shift precision as a function of the number of detected photons at varying intensity noise (reported as standard deviation). **f**, Shift precision as a function of the number of sampling points at varying number of detected photons (see Methods for details). The precision is calculated as the standard deviation of 100 replicates. The reported number of photoelectrons corresponds to the maximum intensity on the photodetector at OPL=0. The star * in the legends indicates the experimental parameters used in our implementation. The red lines in panels a, b and f are linear fit (in the log-log plots) to the right half of the simulation curves.

*Mechanical imaging in biological samples*

After characterizing the optical performance, we proceeded to test the capabilities of our FTBM to image real-world biological samples. For this we acquired FTBM images over a FOV of 245 x 240 x 151 μm in the tail region of a live zebrafish larvae at 2 days post-fertilization (2 dpf). With 56mW of illumination power, corresponding to 0.7μW/pixel, we obtained high signal-to-noise Brillouin shift images of the region surrounding the notochord, from which known anatomical landmarks such as muscle tissue, vacuolated cells and the central canal can clearly be discerned. While small artefacts become present on the far side of the tissue due to a less confined illumination laser there and possible refraction, this nevertheless confirms the in-principle suitability of FTBM for biological imaging applications. Specifically, we highlight that the required illumination energy to reconstruct the Brillouin spectrum in FTBM is only 11µJ/pixel, which compares very favorably to confocal (~0.5-5mJ/pixel), and is similar to line-scan (~10µJ/pixel), BM implementations that were used to acquire similar biological samples. However, we further note that the actual power density per pixel/spectrum are only 0.7µW/pixel for FTBM, but 5-50mW/pixel for confocal, and 0.1mW/pixel for line-scan BM. Since both power density and total energy need to be limited to avoid detrimental effects of photodamage, FTBM will be specifically advantageous for bioimaging applications of photosensitive samples.

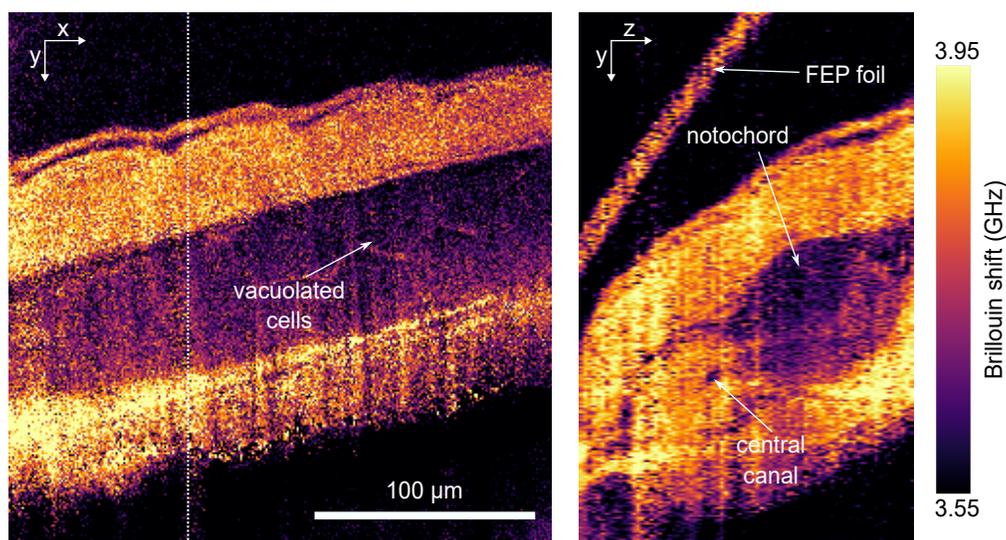

**Figure 4: FTBM imaging of live zebrafish notochord.** The white dotted line indicates the position of the axial cross section shown on the right. Anatomical features of the notochord region such as vacuolated cells and the central canal are visible in the lateral as well as axial views. The image was acquired 2 days post-fertilization. The total illumination power was 56mW, 76 planes were acquired for a total FOV of 245x240x151µm, corresponding to 279x273x75 pixels.

**Discussion**

To summarize, in this work we demonstrated a new approach and key working principle for fast, high-resolution 'full-field' 3D imaging of mechanical properties based on a Fourier-Transform Brillouin spectrometer. We show its ability to visualize mechanical properties of simple phantoms as well as real-world biological samples at substantially higher speed and overall throughput as compared with alternative Brillouin imaging modalities[7–12]. In particular, compared with line-scan Brillouin microscopy, FTBM achieves a 3-fold improvement in terms of effective imaging speed, and a >100-fold reduction in illumination density. Here, in terms of effective pixel time, i.e. number of pixel/spectra measured per second, FTBM achieves full spectral acquisition in ~25µs/pixel, thereby approaching for the first time the performance of fluorescence-based bio-imaging modalities. This substantial overall boost in performance is due to the highly multiplexed measurements afforded by the full-field detection, and could be even further improved by exploiting the full chip of a large sensor sCMOS cameras. Additionally, our FTBM approach can also achieve high spectral precision with consumer grade instrumentation, e.g. higher read-noise cameras (due to the Fellgett's advantage) or less precise translation stages (**Fig. 3a,b**). We expect this will facilitate further uptake by the bioimaging community and aid in overall dissemination. Another distinct advantage of our FT-based approach is the fact that it enables straightforward modification of the spectral frequency range to be sensed while retaining the spectral resolution, which can be achieved through simple optimization of the interferogram sampling. This is unlike common Brillouin spectrometers based on e.g. VIPAs, where larger spectral ranges (FSRs) typically come at the expense of lower spectral resolution. Indeed, in our work we have used different interferogram sampling settings for water and heterogeneous samples (see Methods). For example, in the case of water, which has a single, well defined Brillouin peak and which is typically used to compare Brillouin spectrometer performances, we could indeed achieve an overall throughput of more than ~40k spectra/sec with a resolution of 1.4GHz (note that in our geometry the Brillouin peak of water has a FWHM of 1.6GHz) over a range of 15GHz.

Current limitations of our approach are shared with other modalities that utilize separate excitation and detection paths, such as light-sheet microscopy, namely that optical sample accessibility from two sides must be ensured, and optical refraction effects between the sample and medium can lead to artefacts. The latter could in principle be mitigated by refractive index matching[28] or adaptive optics methods[29]. Further thinning of the light-sheet illumination (currently ~10μm) to match it to the axial resolution of the objective, as well as general optimization of the FT spectrometer's optical design, are also expected to improve overall spatial resolution and image quality. At present, a further practical limitation is the maximum available laser power (<70mW) which limits the power density at sample plane, due to the width of the light-sheet, and generally, of the large FOV afforded by our FTBM. Furthermore, we note that ASE filtering of the illumination laser is critical to obtain high signal-to-noise spectral recordings. Here, we highlight that while elastic (Rayleigh) background scattered light can be effectively (>80dB) suppressed by the Rb cell, the ASE overlaps spectrally with the Brillouin frequency of interest and therefore needs to be adequately filtered out in the illumination arm before reaching the sample.

Finally, we highlight that our approach to undersampling in FT-based spectral imaging is in principle of general nature and can be applied to any symmetric spectrum, although we only confirmed it here for symmetric Brillouin spectra. Future work will focus on its extension to more arbitrary spectra. For example, the same setup with appropriate filters should also be able to acquire concurrent Raman spectra together with the Brillouin signal[26,27], which can be used to correlate the mechanical properties with the local chemical composition of biological specimens, in order to obtain a deeper understanding of the measured mechanics. Going forward, we expect our method to find numerous applications in fast and/or high-throughput Brillouin scattering based imaging applications in and outside biology

**Acknowledgements**

We would like to acknowledge support by the EMBL Heidelberg mechanical and electronic workshops. We are further grateful for useful suggestions from Ling Wang during the initial phase of our project, and to Camilla Autorino and Mantha Lamprousi for providing the zebrafish sample shown in Fig. 4 and the required reagents. R.P. acknowledges support of an ERC Consolidator Grant (no. 864027, Brillouin4Life), the German Center for Lung Research (DZL), as well as research funding 'Life Science' of the Molit Institute. This work was supported by funds from the European Molecular Biology Laboratory.

**Contributions**

C.B. conceived the idea and conceptualized the approach together with R.P. C.B. performed experiments and simulations, and analyzed the data with input from R.P. C.B. and R.P. wrote the manuscript. R.P. acquired funding and supervised the project.

**Disclosures**

The authors have filed a patent application (European Patent EP24163369) related to the methods described in this manuscript.

**Code availability**

The code underlying the simulations presented in this paper are available at https://github.com/prevedel-lab/FT-Brillouin-microscopy

**Data availability**

The raw data underlying the results presented in this paper are not publicly available at this time but will be made public at time of publication.

## Methods

**FTBM setup and design**

The Brillouin illumination and objective configuration is similar to Ref.[10], consisting of two identical objectives (Nikon, ×40, 0.8 numerical aperture, MRD07420, water immersion) mounted in an inverted V geometry. To generate the lightsheet, a plano-convex lens (354330-B, Thorlabs, f=3.1mm) is used to collimate the laser out of a polarization-maintaining fiber (P3-780PM-FC-2, Thorlabs) to a beam diameter of 0.48 mm $1/e^2$ (theoretical). A cylindrical lens (LJ1310L1-B, Thorlabs, f=4.01mm) focuses the light on a plane conjugated to the back focal plane of one of the two objectives. The resulting lightsheet has a theoretical width of 598μm ($1/e^2$), length of 287μm (Rayleigh range) and thickness of 10μm ($1/e^2$).

On the detection side, a plano-convex lens (LA1256-B, Thorlabs, f=300mm) generates an intermediate image (with an effective magnification of 60x). Here a D-shaped mirror it is used to introduce a reference beam at the edge of the FOV. The reference, taken out of the 1% port of a fiber beam splitter (PN780R1A2, Thorlabs), goes through a fiber AOM (Brimrose, TEM-250-50-10-780-2FP), which allows to adjust the intensity of the light and introduces a frequency shift of 250 MHz, to reduce the absorption from the Rb cell, thus avoid the complete suppression of the main laser line (note that the change in wavelength is sufficiently small to not cause a significant phase-shift over the full travel range of the stage). Two plano-convex lenses (cat. no. 39-150, Edmund, f=30mm and cat. no. 39-152, Edmund, f=100mm) collimate and focus the light, respectively, on the D-shaped mirror in the intermediate image plane. Neutral density filters, with a total OD 7, reduce the intensity of the laser so that it can be acquired at the same time of the Brillouin signal without saturating the camera.

A plano-convex lens (LA1979-B-ML, Thorlabs, f=200mm) re-collimates the light from the intermediate image plane, bringing it back to infinity space. There a 75mm long $^{87}$Rb cell (SC-RB87-(25×75-Q)-AR, Photonics Technologies) absorbs the elastically scattered light while transmitting the Brillouin signal (for the beads phantoms and zebrafish sample a 150mm long cell was used instead). A beam expander composed of a f=76mm (cat. no. 49-794, Edmund) and a f=400mm (LA1725-B-ML, Thorlabs) lens expand the beam to 28mm, thus reducing the angles inside the Michelson interferometer.

The Michelson consists of a 50:50 beam splitter (cat. no. 47-572, Edmund), two retroreflectors (HM-15-1E, PLX), one of which is mounted on a long travel range motorized stage (CLL42, SmarAct). After the Michelson a tube lens (f=250mm) forms the image of the sample on a sCMOS camera (C11440-22CU, Hamamatsu). A band pass filter (FBH780-10, Thorlabs) in front of the camera suppresses any unwanted background stray light.

**Imaging**

For the imaging of phantom in Fig. 2, the total power on the sample is 56mW, the exposure time is 100ms. We sampled the interferogram with 5 fine steps with $\Delta L$=97.5nm (OPL=195nm) and 31 coarse steps of $\Delta L$=4mm (OPL=8mm), corresponding to an accessible frequency range of 0-18.75GHz with a resolution of 1.2GHz. The parameters are identical for the zebrafish image in Fig. 4, with 2x binning before data reconstruction.

For the 5 experimental datapoints in Fig. 3 (and Extended Data Fig. 3) the total optical power on the sample (water) and the exposure time per stage position are (33mW, 100ms), (70mW, 100ms), (136mW, 100ms), (168mW, 150ms), (240mW, 300ms). To sample the interferogram, 5 fine steps of $\Delta L$=97.5nm (OPL=195nm) and 21 coarse steps of $\Delta L$=5mm (OPL=10mm) were taken, corresponding to an accessible frequency range of 0-15GHz with a resolution of 1.4GHz. The precision is determined by acquiring 15 images of water, calculating the standard deviation and taking an average of a central region of the resulting image.

**Numerical simulations**

The numerical simulation was performed in Python 3.11.7, numpy 1.26.4, scipy 1.11.4. A theoretical Brillouin spectrum (Lorentzian lineshape) was Fourier-transformed to obtain the raw interferogram, and the following noise sources were considered: Shot noise was added to the simulated interferogram data by drawing each individual sample from a Poisson distribution having an expected value given by $N_{detect}$. Camera noise was subsequently added by drawing samples from a normal distribution with width reported in the legend of Fig. 3a-b. The stage precision in Fig. 3c was taken into account by drawing the stage position from a normal distribution with width reported in the legend and represents typically achievable values. The intensity noise in Fig. 3e was added before the shot noise by drawing samples from a normal distribution with the sigma given by the percentage reported in the legend multiplied by the intensity. The sampling points reported on the x-axis of Fig. 3f corresponds to $N_S$ while keeping the total OPL ($2N_S \cdot \delta\Delta L_n$) constant at 200mm. We also note that, while we assumed a Lorentzian lineshape, experimental spectra can in principle obtain more complex lineshapes, due to the NA broadening expected for our orthogonal geometry[30], which however are not expected to change the main trends reported in Fig. 3.

**Sample preparation**

The phantom is prepared by mixing oil (Immersol W2010, Zeiss) with 0.8% agar (with fluorescein to make it fluorescence) and keeping it in a warm ultrasound water bath during polymerization, so that some beads are formed at the oil agar interface.

A zebrafish larva (2 dpf) from a citrine endogenously labelled alpha-catenin line Gt(ctnna-citrine)ct3a was used in Fig. 4. 1-Phenyl 2-thiourea was added at 0.003% concentration shortly after fertilization to avoid pigmentation. The fish was placed inside the imaging chamber in E3 medium. The sample is separated from the immersion liquid of the objectives (water) by a 12.7µm thick FEP foil (more details can be found in Ref.[10]).

Animal work in this research was carried out at the European Molecular Biology Laboratory (EMBL). All animal care and procedures performed in this study conformed to the EMBL Guidelines for the Use of Animals in Experiments and were reviewed and approved by the Institutional Animal Care and Use Committee.

## Supplementary Figures

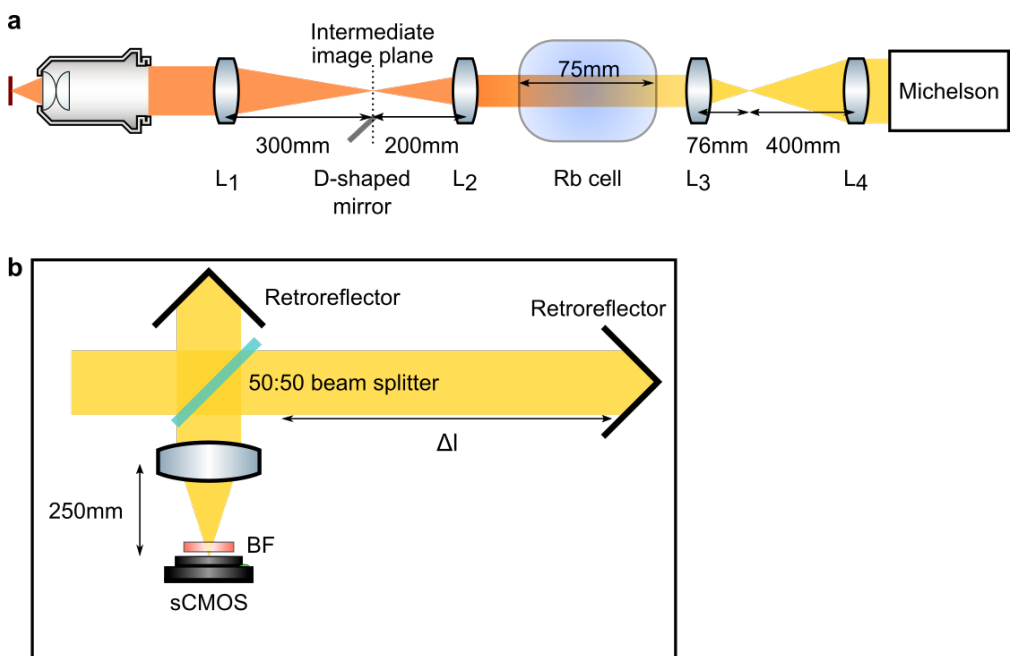

**Extended Data Figure 1: Optical design of the FTBM. a**, Conceptual schematic of the optical layout. An objective (not shown) generates a light sheet (red line on the left). The scattered light is imaged into an intermediate image plane by an identical objective (40x, 0.8NA, Nikon) and a 300mm plano-convex lens. A D-shaped mirror in the intermediate image plane is used to introduce a reference beam in the same optical path. A $^{87}$Rb vapor cell is used to filter the unwanted elastically scattered light. A combination of lenses subsequently prepares the light to be input into the Michelson interferometer (see Methods for more details). **b**, Optical schematic of the Michelson interferometer consisting of a 50:50 beam splitter and 2 hollow retroreflectors. A 250mm tube lens generates the image of the sample on the camera. BF, bandpass filter to suppress stray light (centered at 780nm with a bandwidth of FWHM=10nm).

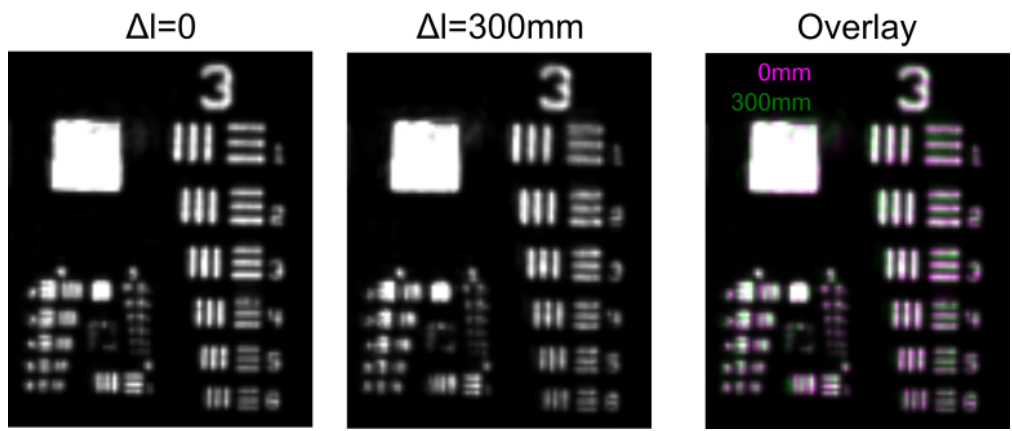

**Extended Data Figure 2: Imaging performances through the FT spectrometer.** Images of a 1951 USAF resolution target, placed in the intermediate image plane, as seen through the FT spectrometer when the fixed arm is blocked and the moving arm is positioned at $\Delta l=0$ and $\Delta l=300$m. We note that group 3 element 6 can still be resolved in both cases, corresponding to a resolution on the sample plane of 0.58μm. The overlay image highlights that, even with an increase of optical path of $2\Delta l=0.6$m, the images are still

overlapping to a high degree which is essential to generate proper interference between the two interferometer arms.

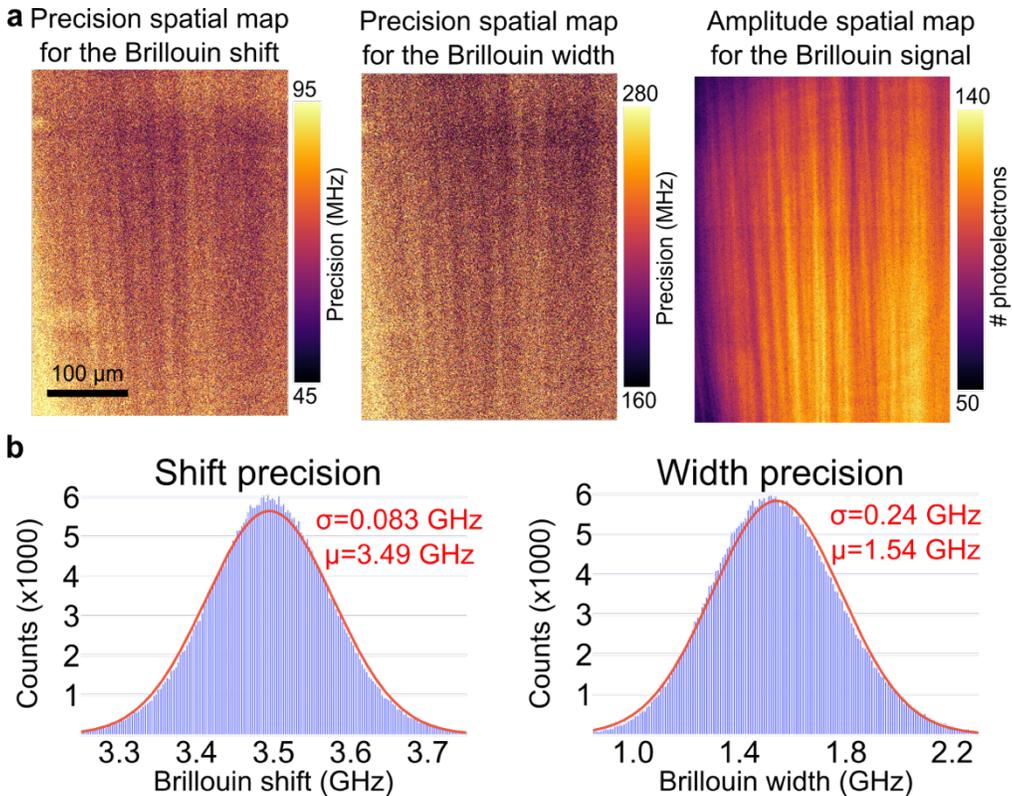

**Extended Data Figure 3: Quantification of the spectral precision of the FTBM in water. a**, Spatial precision map for the Brillouin shift and width of distilled water (553x745 px$^2$). The value (precision) of each pixel is calculated as the standard deviation of 15 replicates (images of water) acquired with the parameters described in the methods (5x20 samples, 100mm total travel range), ~70mW optical power on the sample and 100ms integration time for each sample of the interferogram (i.e. 10s of cumulative exposure time for each image). **b**, Histogram of the Brillouin shift and width for an entire single frame in panel a,b. The solid red curve represents a Gaussian with the mean and standard deviation calculated from the data as indicated.

# Supplementary Notes

## Supplementary Note 1: Definition of arbitrary and typical Brillouin spectra

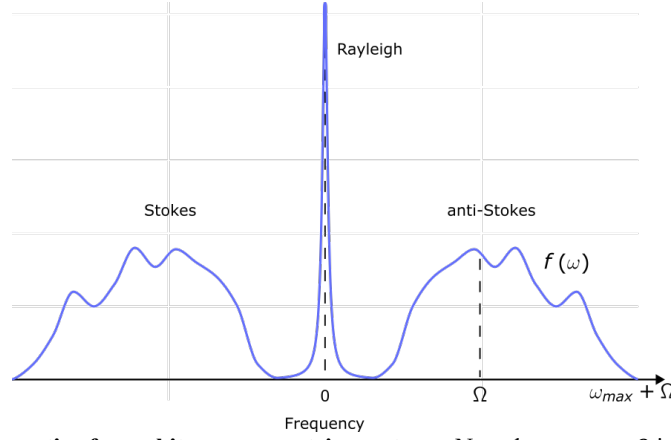

**SI Figure 1: Schematic of an arbitrary symmetric spectrum.** Note that $\omega_{max} + \Omega$ is the maximum permissible frequency.

For simplicity we assume that the optical spectrum has the general form of two symmetric sidebands (labelled as Stokes and anti-Stokes) on the side of a δ-like laser centered at the optical frequency $\omega_L$ (SI Fig. 1). Let's set $f(\omega) \in \mathbb{R}$ to be the anti-Stokes component of the spectrum, shifted by a frequency $\Omega$ from the laser line, and let's introduce the symmetric function $F(\omega) = f(\omega - \Omega) + f(-\omega - \Omega)$ which includes both the Stokes and anti-Stokes components. We can thus write the optical power spectrum as:

$$PS(\omega) = I_L \delta(\omega - \omega_L) + F(\omega - \omega_L) \quad (S1.1)$$

where $I_L$ is the intensity of the laser and $\delta$ is the Dirac-delta.
According to the Wiener–Khinchin theorem:

$$\langle E(t)E(t-\tau)\rangle_t = Re\{\mathcal{F}\{PS(\omega)\}\} \quad (S1.2)$$

where $Re\{z\}$ is the real part of the complex number $z$ and $\mathcal{F}\{g(\omega)\} = \tilde{g}(\tau)$ is the Fourier transform of a generic function $g(\omega)$.
If we introduce $(S1.1)$ in $(S1.2)$ we get:

$$\langle E(t)E(t-\tau)\rangle_t = \cos(\omega_L \tau)\left\{I_L + 2Re\{e^{-i\Omega\tau}\tilde{f}(\tau)\}\right\} \quad (S1.3)$$

We note that the information about $F(\omega)$ is fully contained in the term

$$A(\tau) := I_L + 2Re\{e^{-i\Omega\tau}\tilde{f}(\tau)\} \quad (S1.4)$$

which is the envelope of the term oscillating at the optical frequency $\cos(\omega_L \tau)$.

In the specific case of a typical Brillouin spectrum consisting of a Lorentzian peak with amplitude $A$, shift $\nu_B = \Omega/2\pi$ and linewidth (FWHM) $\Delta\nu_B$, $f(\omega) = \dfrac{A}{1+\left(\dfrac{\omega/\pi}{\Delta\nu_B}\right)^2}$ , $\tilde{f}(\tau) = Ae^{-2|\tau|\cdot\Delta\nu_B}$ and eq. $(S1.4)$ becomes:

$$A(\tau) := I_L + 2Ae^{-2|\tau|\cdot\Delta\nu_B}\cos(\Omega\tau) \quad (S1.5)$$

## Supplementary Note 2: Determination of the sign for the amplitude

From Eq. (S1.4) it follows that $A(\tau)$ can be negative when the optical power of the laser $I_L$ is smaller than the power of the signal of interest (which is desirable). The knowledge of the sign is essential for the proper reconstruction of the spectrum. Experimentally, the effect of a negative sign of $A(\tau)$ is a phase shift of $\pi$ in the $\cos(\omega_L \tau)$ term of the interferogram (Eq. (S1.3)). Therefore, our strategy is to compute the difference between the phase expected in case of monochromatic light and the measured phase; if the absolute value of the difference is greater than $\pi/2$ the sign is considered to be negative. But how to determine the theoretical phase? In principle one could calculate it from the wavelength and the position of the stage. The wavelength is known with extremely high accuracy (<0.1pm), since the laser is locked to the Rb reference, but the accuracy on the stage position needs to be ~10nm, which is challenging to obtain over the full travel range (>100mm). To lessen the requirements on the stage precision/accuracy, we use an internal reference: we introduce a small portion of the main laser at the edge of the FOV in the intermediate image plane (see Extended Data Fig. 1) Such reference beam has a fixed phase relationship with the other spatial points, which is calculated in Supplementary Note 3 and can be experimentally characterized by acquiring the interferogram from a highly scattering sample (or by removing the Rb cell). Note that the characterization needs to be performed only once, if the alignment of the interferometer is not changed. The difference between the phase of each spatial point and the reference beam can then be compared with the "ideal" phase. SI Fig. 2 visually summarizes the described process.

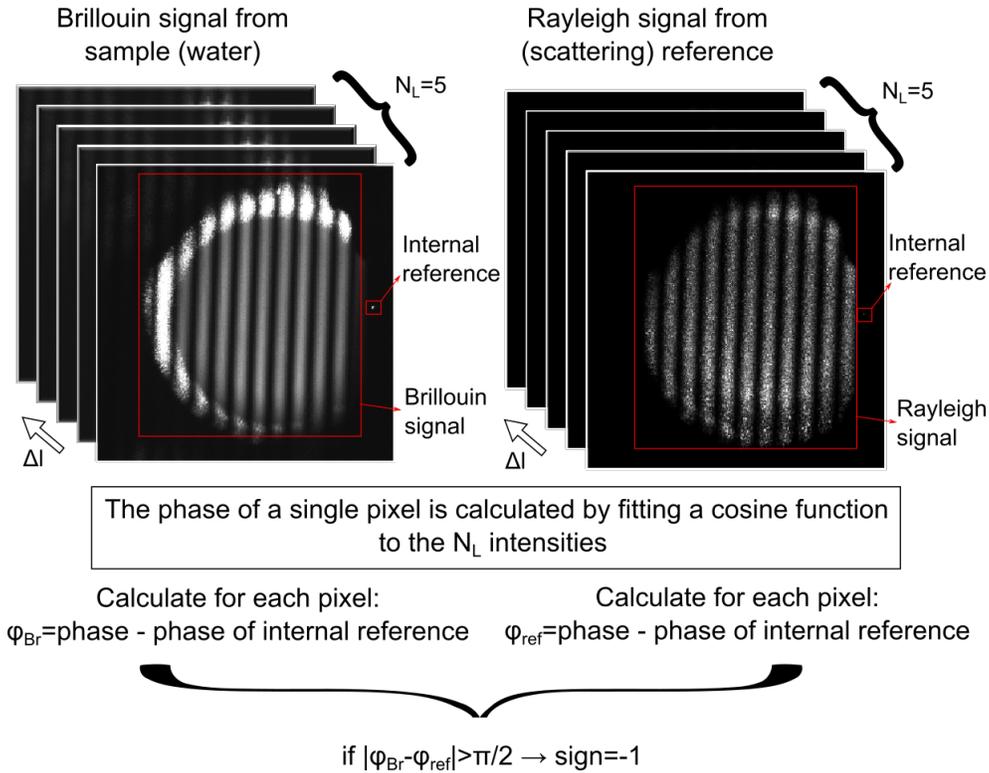

**SI Figure 2: Summary of the procedure used to reconstruct the phase from the raw interferogram data.** See Supplementary Note 2 for details.

**Supplementary Note 3: Phase profile and reference inside the FT-interferometer**

To aid in the optical design of the FT imaging spectrometer, it is useful to calculate the spatial map of the phase as a function of the optical path difference. Since the Michelson is in the infinity space, each lateral position in the sample corresponds to a collimated beam with a different angle with respect to the optical axis, thus undergoing a difference optical path inside the interferometer. Below we derive an expression for the optical path difference (OPD) between the two arms (SI Figure 3) as a function of the relative axial displacement between the retroreflectors $\Delta L$, the relative lateral displacement of the retroreflectors due to not perfect alignment $\delta$ and the angle between the beam and the optical axis $\theta$.

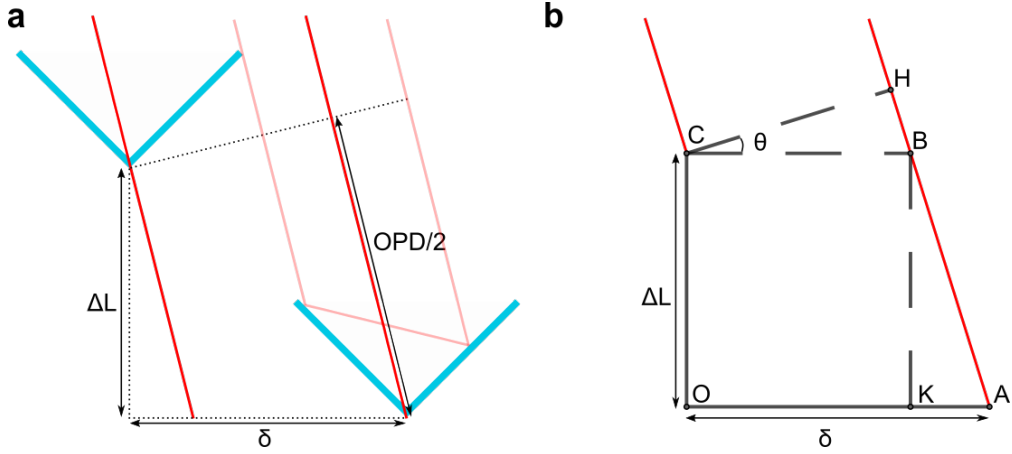

**SI Figure 3: Schematic of the optical path of the rays inside the interferometer. a**, Schematic of the two retroreflectors (light blue) and optical rays going through the vertex (dark red), with the relevant distances are defined. **b**, same as panel a, where the retroreflectors are not shown and auxiliary lines and labels are introduced to aid in the calculations.

The optical path for all the rays within the beam is the same as the one for the ray going through the vertex of the retroreflector.

All the following calculations are performed in the plane $\Sigma$ defined by the optical axis and the ray going through the vertex of the first retroreflector. Under this assumption, $\delta$ correspond to the projection of the physical displacement $\delta_{TOT} = (\delta_x, \delta_y)$ on $\Sigma$, i.e. $\delta = \delta_{TOT} \cdot \cos \Phi$, where $\Phi$ is the angle between the $\Sigma$ and the horizontal (x) plane.

$$\overline{CB} = \overline{OA} - \overline{KA} = \delta - \Delta L \tan \theta$$

$$\overline{BH} = \overline{CB} \sin \theta = (\delta - \Delta L \tan \theta) \sin \theta$$

$$\overline{AB} = \frac{\overline{BK}}{\cos \theta} = \frac{\Delta L}{\cos \theta}$$

$$\overline{AH} = \overline{AB} + \overline{BH} = \Delta L \left( \frac{1}{\cos \theta} - \frac{\sin^2 \theta}{\cos \theta} \right) + \delta \sin \theta = \Delta L \cos \theta + \delta \sin \theta$$

In the approximation of $\theta \ll 1$ ($\theta < 0.02$ rad in our experimental realization):

$$\overline{AH} \approx \Delta L \left( 1 - \frac{\theta^2}{2} \right) + \delta \cdot \theta$$

The angle $\theta$ corresponds to the spatial position on the camera $(x_c, y_c)$ (where the intersection of the optical axis with the sensor is set to (0,0)) through the relationship $\theta \approx \frac{r}{f}$, where $r := \sqrt{x_c^2 + y_c^2}$ and $f$ is the focal length of the tube lens. Therefore, in terms of the coordinates of the camera:

$$\overline{AH} \approx \Delta L \left(1 - \frac{r^2}{2f^2}\right) + \delta \frac{r}{f} = -\frac{\Delta L}{2f^2} r^2 + \frac{\delta}{f} r + \Delta L$$

Since $\delta = \delta_{TOT} \cdot \cos \Phi$, $\delta \cdot r$ is the projection of $\delta_{TOT}$ on the direction of $r$ and it can thus be written as the inner product $\delta_x \cdot x_c + \delta_y \cdot y_c$.

Finally, the OPD is:

$$OPD = 2\overline{AH} = -\frac{\Delta L}{f^2}(x_c^2 + y_c^2) + 2\frac{\delta_x \cdot x_c + \delta_y \cdot y_c}{f} + 2\Delta L$$

Given a fixed position of the retroreflector ($\Delta L$=const), each spatial position on the camera corresponds to a beam that underwent a different optical path inside the interferometer, thus generating fringes. The spatial map of the phase of the fringes is given by:

$$\varphi(x_c, y_c) = \varphi_0 + k \cdot OPD = \varphi'_0 + \frac{2\pi}{\lambda f}\left[-\frac{\Delta L}{f}(x_c^2 + y_c^2) + 2(\delta_x \cdot x_c + \delta_y \cdot y_c)\right]$$

The period of the fringes $\Lambda$ (i.e. the distance over which the phase changes by $2\pi$) is a linear function of $x_c$ and $y_c$, since $\varphi(x_c, y_c)$ is quadratic in $x_c, y_c$. For simplicity we calculate $\Lambda$ only for the x direction, since the expression for the y direction is equivalent. We can calculate it by imposing that the $|\varphi(x_{c,2}) - \varphi(x_{c,1})| = 2\pi$ and assuming $x_c \approx \frac{x_{c,1} + x_{c,2}}{2}$

$$\Lambda = \frac{\lambda f^2}{2|-\Delta L \cdot x_C + f \cdot \delta_x|}$$

We must make sure that the pixel size on the camera $\Delta x_C$ is sufficiently small to sample the fringes properly, i.e. $\Delta x_C < \Lambda/4$. For the following calculation we assume $\delta_x = 0$ (i.e. perfect alignment).

$$\Delta x_C < \frac{\Lambda}{4} = \frac{\lambda f^2}{8\Delta L \cdot x_C} < \frac{\lambda f^2}{8\Delta L_{max} \cdot x_{C,max}}$$

$x_{C,max} = M_{eff} \cdot FOV/2$ where $M_{eff}$ is the overall magnification from the sample plane to the camera and $FOV$ is the field of view on the sample plane.

$$\Delta x_C < \frac{\lambda f^2}{4\Delta L_{max} \cdot M_{eff} \cdot FOV}$$

In our design $M_{eff} = 14.25$, $FOV \approx 350$mm and $\Delta L_{max} = 100$mm, thus $\Delta x_C < 25$mm, which is compatible with the pixel size of our camera (6.5mm).